\documentstyle[sprocl,epsfig]{article}

\begin{document}

\title{A CONCEPTION OF THE PHOTON COLLIDER BEAM DUMP }

\author{L.I.~SHEKHTMAN, V.I.~TELNOV~\footnote{Speaker}}

\address{Institute of Nuclear Physics,  630090, Novosibirsk, Russia}


\newcommand{\M}{\mbox{m}}
\newcommand{\n}{\mbox{$n_f$}}
\newcommand{\EP}{\mbox{e$^+$}}
\newcommand{\EM}{\mbox{e$^-$}}
\newcommand{\EPEM}{\mbox{e$^+$e$^-$}}
\newcommand{\EMEM}{\mbox{e$^-$e$^-$}}
\newcommand{\GG}{\mbox{$\gamma\gamma$}}
\newcommand{\GE}{\mbox{$\gamma$e}}
\newcommand{\GP}{\mbox{$\gamma$e$^+$}}
\newcommand{\TEV}{\mbox{TeV}}
\newcommand{\GEV}{\mbox{GeV}}
\newcommand{\LGG}{\mbox{$L_{\gamma\gamma}$}}
\newcommand{\LGE}{\mbox{$L_{\gamma e}$}}
\newcommand{\LEE}{\mbox{$L_{ee}$}}
\newcommand{\WGG}{\mbox{$W_{\gamma\gamma}$}}
\newcommand{\EV}{\mbox{eV}}
\newcommand{\CM}{\mbox{cm}}
\newcommand{\MM}{\mbox{mm}}
\newcommand{\NM}{\mbox{nm}}
\newcommand{\MKM}{\mbox{$\mu$m}}
\newcommand{\SEC}{\mbox{s}}
\newcommand{\CMS}{\mbox{cm$^{-2}$s$^{-1}$}}
\newcommand{\MRAD}{\mbox{mrad}}
\newcommand{\IND}{\hspace*{\parindent}}
\newcommand{\E}{\mbox{$\epsilon$}}
\newcommand{\EN}{\mbox{$\epsilon_n$}}
\newcommand{\EI}{\mbox{$\epsilon_i$}}
\newcommand{\ENI}{\mbox{$\epsilon_{ni}$}}
\newcommand{\ENX}{\mbox{$\epsilon_{nx}$}}
\newcommand{\ENY}{\mbox{$\epsilon_{ny}$}}
\newcommand{\EX}{\mbox{$\epsilon_x$}}
\newcommand{\EY}{\mbox{$\epsilon_y$}}
\newcommand{\BI}{\mbox{$\beta_i$}}
\newcommand{\BX}{\mbox{$\beta_x$}}
\newcommand{\BY}{\mbox{$\beta_y$}}
\newcommand{\SX}{\mbox{$\sigma_x$}}
\newcommand{\SY}{\mbox{$\sigma_y$}}
\newcommand{\SZ}{\mbox{$\sigma_z$}}  

\maketitle\abstracts{Photon beams at photon colliders are very narrow,
powerful and can not be deflected. For the beam dump at the TESLA-like
collider we suggest to use a long gas (Ar) spoiler in front of the
water absorber, this solves the overheating and mechanical stress
problems. The neutron background at the interaction point is
estimated.}
\vspace{-0.5cm}
The beam dump at the linear collider TESLA suggested for the \EPEM\
interaction region consists of the deflecting magnets (deflect bunches
inside one train) and the water vessel at the distance 250 m from the
interaction point.~\cite{TESLATDR} However it does not suit for the
photon collider because the photon beam is neutral. Characteristic
beam parameters: $N=2\cdot 10^{10}$, the number of bunches in one
train 2820, $\Delta t = 337$ nsec, $\nu = 5$ Hz, the angular
divergence $\sigma_{\theta_x} \sim 3 \cdot 10^{-5}$, $\sigma_{\theta_y}
\sim 10^{-5}$. The beam is mixed, about half of the energy is carried
by electrons and half by photons. 

  Our scheme of the beam dump for the photon collider is depicted in
Fig.1.  In addition to the TESLA TDR solution we added the beam
spoiler of 4--5 radiation length thickness, the cheapest solution is
Ar at the pressure 3--5 atm ($X_{Ar}=110$ m at P=1 atm). The
deflecting (rotating) magnets are needed to decrease the stress in the
entrance Al-Be window situated at the distance 100 m from the IP. The
circle radius of $R=$ 0.5--1 cm is sufficient. The thickness of the
entrance window is small, therefore the density of electrons from the
pair production by the narrow photon beam in the entrance window is
acceptable. In the gas, photons and electrons produce showers and due
to the multiple scattering the density of particle at the exit window
of the gas vessel and entrance window of the water beam dump is also
acceptable. A third critical place is the most dense point in the
shower situated in the water beam dump between the entrance window and
the shower maximum. In the chosen scheme the rise of the water
temperature at this point is also acceptable.
\begin{figure}[ht] 
     \begin{center}
     \vspace*{-1.cm}
   \epsfig{file=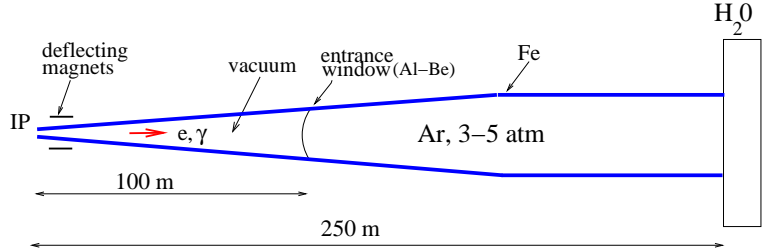,width=10cm}
     \end{center}
     \caption{The scheme of the beam dump for TESLA.} 
     \end{figure}

\begin{figure}[ht] 
     \begin{center}
     \vspace*{-0.2cm}
   \epsfig{file=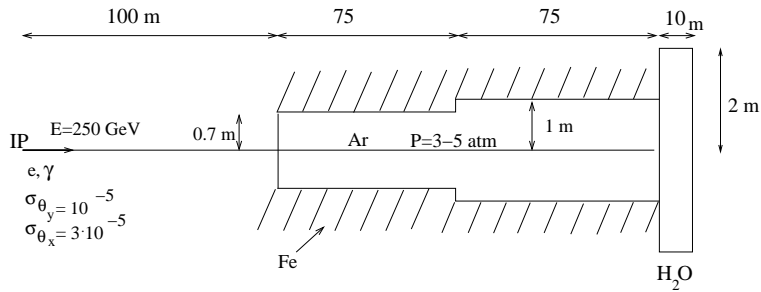,width=10cm}
    \vspace*{-0.3cm}
     \end{center}
     \caption{The scheme used in the simulation.} 
   \vspace*{-0.5cm}
   \end{figure}
 The simulation using FLUKA code was done for the geometry shown in
   Fig.2. Some preliminary results are the following.  The maximum
   $\Delta T$ at the entrance Be-Al window is about 40$^\circ$ for
   R=0.5 cm (sweeping radius). For the removal of the heat the thermal
   conductivity is sufficient, the gas cooling can be added, if
   necessary.  The maximum local $\Delta T$ at the exit Be-Al window
   is small, about $10^\circ$.  The maximum local $\Delta T$ in the
   water dump after passage of the train from 250 GeV photons is 75,
   50, 25$^\circ$ at the Ar pressure 3, 4, 5 atm, respectively, and by
   a factor of 2 lower for incident electrons.

The problem of the mechanical stress in solid materials in the TESLA
beam dump is not important because the train duration is much longer
than the decay time of local stress ($r/v_{sound} \sim 1$ $\mu sec$).
It is more serious for warm-LC with short trains.

The simulation gives also an estimate of the neutron flux at IP.  For
$10^5$ incident 250 GeV electrons and $P_{Ar}=4$ atm there are 6
neutrons at the IP plain $z=0$ with the radial coordinates $r=$ 1.5,
2.5, 4.5, 14.5, 18.5 21.5 m.  Due to the collimation by the
Fe tube we do not expect the uniform density, the density per cm$^2$
should be larger near the axis. Assuming the uniform density for three
neutrons closest to the axis we find the flux $5\cdot 10^{-11}$
n/cm$^2$ per incident electron or about $1.5\cdot 10^{11}$ n/cm$^2$
for $10^{7}$ sec run time.

   It is remarkable that after the replacement of the first 20 m of Ar by
   $H_2$ at the same pressure there is only one neutron at $r=1.5$ m
   for $8\cdot 10^5$ incident electrons. With account of collimation
   by the tube it means the decrease of the neutron flux at least by a
   factor of ten!

\end{document}